\documentclass{ws-procs975x65}
\usepackage{mathrsfs}
\usepackage{psfrag}

\DeclareSymbolFontAlphabet{\mathrsfs}{rsfs}
\DeclareMathAlphabet{\mathcal}{OMS}{cmsy}{m}{n}
\newcommand{\scri}{\mathrsfs{I}}

\newcommand{\be}{\begin{equation}}
\newcommand{\ee}{\end{equation}}

\begin{document}

\title{HYPERBOLOIDAL FOLIATIONS WITH $\scri$-FIXING IN SPHERICAL SYMMETRY}

\author{ANIL ZENGINO\u{G}LU$^1$, SASCHA HUSA$^2$}

\address{$^1$Max-Planck-Institut f\"ur Gravitationsphysik (AEI)\\ 
D-14476 Potsdam, Germany\\ \email{anil@aei.mpg.de}}
\address{$^2$ Universit\"at Jena, Theoretisch-Physikalisches Institut,\\
Max-Wien-Platz 1, 07743 Jena, Germany\\
\email{sascha.husa@uni-jena.de}}

\begin{abstract}
We study in spherical symmetry the conformal compactification for hyperboloidal foliations with nonvanishing constant mean curvature. 
The conformal factor and the coordinates are chosen such that null infinity is at a fixed radial coordinate location.
\end{abstract}

\bodymatter

\section*{}

Hyperboloidal surfaces in asymptotically flat spacetimes have first been used for an initial value formulation of the Einstein equations by Friedrich\cite{Friedrich83}. 
Instead of approaching spatial infinity as Cauchy surfaces do, they reach null infinity, $\scri$, which makes them suitable for radiation extraction. 
Contrary to characteristic surfaces, these spacelike surfaces are as flexible as Cauchy surfaces and they can be used in numerical calculations with the 3+1 approach based 
on a hyperboloidal initial value problem\cite{Frauendiener04,Huebner01,Husa02,Husa03}. 
We want to study the conformal compactification\cite{Penrose63} of hyperboloidal foliations in spherical symmetry. 
It has been suggested\cite{Andersson02,Frauendiener98,Husa05} that conformal compactifications in which $\scri$ is kept at a fixed spatial coordinate location 
might be useful for testing new ideas in numerical calculations. 
Here we explicitly discuss the simplest cases, namely the Minkowski and Schwarzschild spacetimes.

The physical line element in spherical symmetry can be written as
\be \label{metric} \tilde{g} = (-\tilde{\alpha}^2 +\tilde{h}^2\tilde{\beta}^2)\,dt^2 + 2 \tilde{h}^2\tilde{\beta}\, dt\,d\tilde{r} + \tilde{h}^2 \,d\tilde{r}^2 + \tilde{r}^2 \,d\sigma^2. \ee
$d\sigma^2$ is the standard metric on $S^2$ and the lapse $\tilde{\alpha}$, the shift $\tilde{\beta}$, and the spatial metric function $\tilde{h}$ are functions of the coordinates $(t,\tilde{r})$ only. We assume that  the metric (\ref{metric}) admits a regular conformal compactification and that the time coordinate $t$ is such that $t$=const.~hypersurfaces are hyperboloidal hypersurfaces. We do not compactify the time direction. The conformal compactification, $g=\Omega^2 \tilde{g}$, can be done such that
\be \label{conf_comp} \Omega^2(\tilde{h}^2d\tilde{r}^2+\tilde{r}^2d\sigma^2)= h^2\,dr^2 + r^2 \,d\sigma^2 \ee
with respect to a compactifying radial coordinate $r$. 
Note that we have some freedom here. One can require for example $h=1$ 
which leads to $r$ being the proper distance, however, then the radial coordinate transformation can not, in general, be written in explicit form\cite{Moncrief06}. 
By keeping $h$, we have the freedom to prescribe the conformal factor in terms of a compactifying radial coordinate $r$ and the coordinate tranformation is explicit. 
The relation (\ref{conf_comp}) implies for a given conformal factor $\Omega(r)$ 
a coordinate transformation $\tilde{r}=\Omega^{-1}\,r$ so that $d\tilde{r}=(\Omega - r\,\Omega')\Omega^{-2}\,dr$. 
Then the spatial metric function transforms as $h = (\Omega - r\,\Omega')\Omega^{-1}\tilde{h}$.
For the regularity of this conformal compactification, $\tilde{h}(t,\tilde{r})$ needs to have a specific asymptotic fall-off behaviour for $\tilde{r}\to\infty$ 
on the hyperboloidal surfaces of constant $t$. 
A simple choice for the conformal factor that we will study is $\Omega=(1-r)$, which implies $h = \Omega^{-1}\tilde{h}$. 
This is not a good choice at the origin, but we are interested in the asymptotic region.

\begin{figure}[ht]
  \centering
  \psfrag{i+}{$i^+$}
  \psfrag{i0}{$i^0$}
  \psfrag{i-}{$i^-$}
  \psfrag{S-}{$\scri^-$}
  \psfrag{S+}{$\scri^+$}
  \psfrag{u}{$V$}
  \psfrag{v}{$U$}
\includegraphics[width=0.9\textwidth]{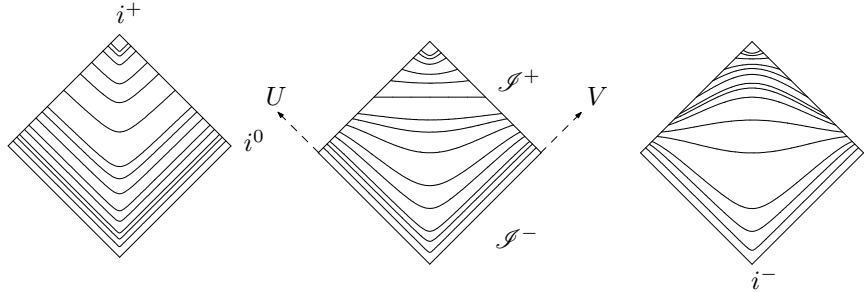}
\caption{\label{cmc_fol} CMC-foliations of the Minkowski spacetime for $K=15,3,1.5$}
\end{figure}

A simple example for hyperboloidal foliations are constant mean curvature (CMC) foliations. We write the line element in Minkowski spacetime with standard coordinates $(\tilde{t},\tilde{r})$ as $\tilde{\eta}=-d\tilde{t}^2+d\tilde{r}^2+\tilde{r}^2\,d\sigma^2$. We introduce $t(\tilde{t},\tilde{r})=\tilde{t}-\sqrt{a^2+\tilde{r}^2}$ as a new time coordinate. The constant $a\in\mathbb{R}$ is related to the constant mean extrinsic curvature $K$ of the level sets of $t(\tilde{t},\tilde{r})$ by $a=3/K$. We use the convention in which positive $K$ means increasing volume to the future. The sign of $K$ determines whether the surfaces reach $\scri^+$ or $\scri^-$. To get a feeling for these surfaces, we analyse them in the familiar compactification of the Minkowski spacetime given by the transformation\cite{Frauendiener04}
\[ \tilde{t}(V,U) = \frac{1}{2} (\tan V + \tan U), \qquad \tilde{r}(V,U) = \frac{1}{2} (\tan V - \tan U). \]
The subsequent rescaling with the conformal factor $\Omega=\cos V \cos U$ leads to 
\[ \eta = \Omega^2 \tilde{\eta} = -dU dV + \frac{\sin^2 (V-U)}{4} d\sigma^2.\]
$\scri^+$ is at $V=\pi/2$ in these coordinates. Embedding our hyperboloidal surfaces into the conformally extended Minkowski spacetime leads to
\[ t(V,U) = \frac{1}{2} (\tan V + \tan U) - \sqrt{a^2 + \frac{1}{4}(\tan V - \tan U)^2}. \]
Writing a series expansion in $\cot V$ near $\scri^+$,
\[ t(V,U) \approx \tan U - a^2 \cot V - a^2 \tan U \cot^2 V + O(\cot^3 V),   \quad \mathrm{for} \quad V\to \pi/2, \]
we see that the cut at $\scri^+$ depends on the value of $t$ via $U(t,V)\big|_{\scri^+}=\arctan t$, but does not depend on the mean extrinsic curvature which determines the angle of the cut. Fig.~\ref{cmc_fol} shows three foliations in the Penrose diagram of Minkowski spacetime for the same set of values of $t$, but different values of $K$. Null surfaces in a Penrose diagram have an angle of 45 degrees to the horizontal. As each plotted surface is spacelike, their angle is smaller.
By the choice of the mean extrinsic curvature, we can control the behaviour of the surfaces in the interior without changing their asymptotics. For example, by choosing $|K|$ small, we can make the hyperboloidal surfaces behave, in a certain sense, more similar to Cauchy slices.

In Penrose diagrams of Minkowski and Schwarzschild, the null generators of $\scri^+$ converge. To avoid the corresponging loss of resolution in numerical calculations, we would like to fix the radial coordinate location of $\scri^+$. The Minkowski metric in the time coordinate $t$ of a CMC-foliation reads
\[ \tilde{\eta} = -dt^2-\frac{2\tilde{r}}{\sqrt{a^2+\tilde{r}^2}}\,dt d\tilde{r} + \frac{a^2}{a^2+\tilde{r}^2}\,d\tilde{r}^2+\tilde{r}^2\,d\sigma^2. \]
Conformal compactification, $\eta=\Omega^2\tilde{\eta}$, using (\ref{conf_comp}) with $\Omega=1-r$ results in 
\be \label{cmc_mink}\eta=-(1-r)^2dt^2 -\frac{2r}{\sqrt{a^2(1-r)^2+r^2}}\,dt dr + \frac{a^2}{a^2(1-r)^2+r^2}\,dr^2+r^2d\sigma^2. \ee

For the Schwarzschild spacetime $\tilde{g}_s$, the general family of spherically symmetric constant mean curvature surfaces has been constructed in \cite{Malec03}. Conformal compactification, $g_s=\Omega^2 \tilde{g}_s$, with $\Omega=1-r$ results in
\be \label{cmc_ss} g_s = -\left(1-\frac{2 m (1-r)}{r}\right)(1-r)^2 dt^2- \frac{2\left(C(1-r)^3-K r^3/3\right)}{P(r)}\,dt dr + \frac{r^4}{P^2(r)}\,dr^2+r^2 d\sigma^2, \ee
where $m, K$, and $C$ are constants and 
\[ P(r):=\left(\left(C(1-r)^3-K r^3/3)\right)^2+\left(1-\frac{2 m (1-r)}{r}\right)(1-r)^2 r^4\right)^{\frac{1}{2}}.\]
As seen in the examples (\ref{cmc_mink},\ref{cmc_ss}), to keep $\scri^+$ at a fixed radial coordinate location we need an inward pointing shift vector in the asymptotic region.
\section*{Acknowledgments}
The authors thank Helmut Friedrich for discussions. This work was supported in part by the SFB/Transregio 7 of the German Science Foundation.

\end{document}